\documentclass[conference]{IEEEtran}
\IEEEoverridecommandlockouts
\usepackage{cite}
\usepackage{amsmath,amssymb,amsfonts}
\usepackage{algorithmic}
\usepackage[ruled,linesnumbered]{algorithm2e}
\usepackage{graphicx}
\usepackage{textcomp}
\usepackage{xcolor}
\def\BibTeX{{\rm B\kern-.05em{\sc i\kern-.025em b}\kern-.08em
    T\kern-.1667em\lower.7ex\hbox{E}\kern-.125emX}}
\begin{document}

\title{Exploring the Equivalence between Dynamic Dataflow Model and Gamma - General Abstract Model for Multiset mAnipulation}

\author{\IEEEauthorblockN{Rui R. Mello Jr}
\IEEEauthorblockA{\textit{Federal University of Rio de Janeiro} \\ \textit{Brazilian Navy Research Institute}\\
Rio de Janeiro, Brazil \\
ruirodrigues@cos.ufrj.br}
\and
\IEEEauthorblockN{Leandro S. Araujo}
\IEEEauthorblockA{\textit{Federal University of Rio de Janeiro} \\
Rio de Janeiro, Brazil \\
lsantiago@cos.ufrj.br}
\and
\IEEEauthorblockN{Tiago A. O. Alves}
\IEEEauthorblockA{\textit{State University of Rio de Janeiro} \\
Rio de Janeiro, Brazil \\
tiago@ime.uerj.br}
\and
\IEEEauthorblockN{\qquad \qquad Leandro A. J. Marzulo}
\IEEEauthorblockA{\textit{\qquad \qquad Google Research} \\
\qquad \qquad Sunny Vale, CA \\
\qquad \qquad leandro.marzulo@gmail.com}
\and
\IEEEauthorblockN{\qquad \qquad Gabriel A. L. Paillard}
\IEEEauthorblockA{\textit{\qquad \qquad Federal University of Cear\'a} \\
\qquad \qquad Cear\'a, Brazil \\
\qquad \qquad gabriel@virtual.ufc.br}
\and
\IEEEauthorblockN{\qquad \qquad Felipe M. G. Fran\c{c}a}
\IEEEauthorblockA{\textit{\qquad \qquad Federal University of Rio de Janeiro} \\
\qquad \qquad Rio de Janeiro, Brazil \\
\qquad \qquad felipe@cos.ufrj.br}
}

\maketitle

\begin{abstract}

With the increase of the search for computational models where the expression of parallelism occurs naturally, some paradigms arise as options for the next generation of computers. In this context, dynamic Dataflow and Gamma --- \textbf{G}eneral \textbf{A}bstract \textbf{M}odel for \textbf{M}ultiset m\textbf{A}nipulation) ---  emerge as interesting computational models choices. In the dynamic Dataflow model, operations are performed as soon as their associated operators are available, without rely on a Program Counter to dictate the execution order of instructions. The Gamma paradigm is based on a parallel multiset rewriting scheme. It provides a non-deterministic execution model inspired by an \textit{abstract chemical machine} metaphor, where operations are formulated as reactions that occur freely among matching elements belonging to the multiset.
In this work, equivalence relations between the dynamic Dataflow and Gamma paradigms are exposed and explored, while methods to convert from Dataflow to Gamma paradigm and vice versa are provided. 
It is shown that vertices and edges of a dynamic Dataflow graph can correspond, respectively, to reactions and multiset elements in the Gamma paradigm. 
Implementation aspects of execution environments that could be mutually beneficial to both models are also discussed. This work provides the scientific community with the possibility of taking profit of both parallel programming models, contributing with a versatility component to researchers and developers. Finally, it is important to state that, to the best of our knowledge, the similarity relations between both dynamic Dataflow and Gamma models presented here have not been reported in any previous work.
\end{abstract}

\begin{IEEEkeywords}
Gamma, Dataflow, Parallel Programming, Computational Models, Equivalence, Similarity.
\end{IEEEkeywords}

\section{Introduction} \label{sec:introduction}

Parallel computing has been widely used as a tool to obtain performance in a scenario where the von Neumann architecture is close to the hardware`s performance exploitation limits \cite{7423190}. In the other words, despite the validity of Moore`s law, possible technological improvements are not being converted into performance in a proportional way. Thus, the use of multi-core processors is currently growing. However, programming in parallel manner may not be a simple task. The multiple execution lines management, identification of parallel parts of code, distribution and control tasks between processors, load control, besides entire problem modeling in order to extract the maximum of parallelism makes writing parallel programs a not trivial task.

In this context, the use of models where the exhibition of parallelism occurs in a natural way has been increasing. Among them, the dataflow model presents a great potential to expose the parallelism. Here, the program execution is driven by the data, in contrast with the von Neumann model where the execution is guided by the control flow (there is a program counter that guides the search of the next instruction to be executed). In this way, the dataflow model can be represented by a directed graph, where the vertices represent the operations to be performed and the edges represent the data (operands) used by these operations. Therefore, once its operands are available, the operation can be performed.

On the other hand, Gamma (General Abstract Model for Multiset mAnipulation) was proposed, in 1986, as a formalism for program specification based on the parallel multiset rewriting. In Gamma the execution model is nondeterministic, since its elements can react freely in a parallel way, making parallelism implementation details transparent to the programmer. The model comprises a single database called multiset, where the data used by computation are represented by multiset elements. This computational model presents a metaphor of chemical reactions, where the multiset (chemical solution) is composed of several elements (molecules). Actions (chemical reactions) are specified to perform on the multiset elements, according to a set of conditions (reaction conditions).

In this paper we explore for the first time the similarity between these two computational models, Gamma and Dataflow. Through empirical and formal tests we show that these models are equivalent, making it possible to transform a dataflow graph into Gamma code and vice versa. This way, it is possible to provide the scientific community with a series of benefits that comprise exploring and analyzing, in a code written in Gamma, speculative and out-of-order dataflow execution \cite{4685743}, performing instructions trace reuse \cite{doi:10.1002/cpe.4937}, among others. In addition, a program initially represented by a dataflow graph can be exploited in an execution environment quite suitable, for instance, to an Internet of Things (IoT) environment.


\subsection{Motivation} \label{subsec:motivation}

The dataflow and Gamma computational models present a surprising similarity. Both models present a natural manner to express the parallelism, making transparent to the developer details related to the parallelism implementation. However, these two paradigms presents substantial differences, with benefits of use in both models. In this way the motivation for this work arises, that consists in to take advantage of the benefits of both models through the proof of similarity between them. Thus, a program written in Gamma could take advantage of the benefits of a series of studies performed on programs expressed through a dataflow graph. Besides, programs in dataflow could be exported to a platform with potential for use in an IoT environment, for instance. in addition, until the present moment, we didn't find any study that explore this similarity. 


\subsection {Objectives} \label{subsec:objectives}

This study has the objective of to exploit the equivalence between two computational models that explore the parallelism in a natural way. Before that, some empirical test was realized and we propose an transformation algorithm to convert a dataflow graph in a code written according to a Gamma formalism.

The rest of the paper is organized as follow: the section \ref{sec:background} present the basics concepts about Gamma and dataflow needed to this paper. In section \ref{sec:similarity_formalization} some examples is showed where empirical tests where realized, the transformation algorithm is presented and its transformations is discussed. Finally, the section \ref{sec:concluding_remarks} present the conclusion and some future works.


\section{Background} \label{sec:background}

The present section addresses the theoretical grounds necessary for the understanding of the Gamma and dataflow computational models.


\subsection {Dataflow Model} \label{subsec:dataflow}

Dataflow model \cite{642111} presents a natural way to express parallelism by describing a program as dataflow graph, where nodes represent tasks or instructions and the edges that connect these nodes indicate their direct data dependencies. The execution of an instruction starts as soon as its input operands are ready. Instructions (or tasks) that are not connected by a path in the graph can run in parallel according to availability of computational resources. The run-time processor does not rely on Program Counter (PC) and the global state, since each operand is directly transferred from producer nodes to consumers nodes.

As dataflow is not dependent on PC, control branches in the program are executed by modifying the flow of data in the dataflow graph at runtime. For example, in a \textit{if-then-else} statement, instructions are grouped in separated subgraphs corresponding to \textit{if} and \textit{else} blocks. After validating the result of logic expression, the operands should be sent to the correct subgraph. To deal with control branches, there is a \textit{steer} node that receives the data operand and a boolean operand, and selects one of the two paths to submit the data operand depending on boolean operand (\textit{true} or \textit{false}). Loops are implemented like control branches by changing the flow of data to some preceding node. Dynamic dataflow enables a loop to run multiple instances of its iterations in parallel. To avoid nodes of current iteration send operands to nodes of past iterations, all operands hold a tag representing an instance number which is always matched before starting the node's execution. Thus, an instruction only runs whether all input operands are ready with the same tag. Management of the loop iterations are supported by a special node \textit{inctag} which increases the tag at beginning of next iteration. Function calling might be supported by manipulating tags following the similar solutions applied in loops \cite{5645392}.

Dataflow runtime systems have emerged as appealing solutions to create parallel programs for multi and many-core environments \cite{1635961,6008964,5284364,Alves:2011:TET:1989576.1989583,doi:10.1142/S0129626411000151,6628287}. Dataflow paradigm is virtually supported on traditional multicore machines, where each core is a virtual Processing Element (PE) that runs the dataflow firing rule. PE dispatches task or function executions whenever incoming operands are available, while the block of code from task are straightforwardly executed on target machine without virtualization. So independent code blocks are triggered by distinct PEs, leveraging parallelism exploitation.


\subsection{Gamma}

 The word GAMMA is an acronym for \textbf{G}eneral \textbf{A}bstract \textbf{M}odel for \textbf{M}ultiset m\textbf{A}nipulation). This paradigm was proposed in 1986 by Ban\^{a}tre and M{\'e}tayer \cite{report-gamma1} and can be defined as a formalism for program specifications based on parallel multiset rewriting. It refers to a nondeterministic execution model, since the multiset elements of can interact freely and naturally parallel way.
 
 The Gamma model is often presented metaphorically as chemical reactions. In this sense, the paradigm has a unique database (the multiset) composed for many elements (molecules) that can be manipulated by several reactions. This reactions correspond to the operations that can be performed over the elements, according to a set of pre-defined conditions (reactions conditions). So, the reactions can run freely over the multiset elements, in a nondeterministic way, and allowing the abstraction of details that make difficult to develop programs in a parallel programming languages.

The execution of a Gamma program occurs through modifying the multiset by exclusion, inclusion and transformation of existing elements, through the performance of pair of functions, composed of conditions/actions over the multiset. Here, the end of the computation occurs when a steady state is reached, where all the reactions have finished their execution, and there are no reaction conditions able to react (global termination state).

According to \cite{artigo-paillard}, we can formally define the $\Gamma$ operator, as follow:

\begin{equation}
\begin{split}	
\Gamma((R_{1}, A_{1}), ... ,(R_{m}, A_{m}))(M)=\\ 
if \hspace{.2cm} \forall \hspace{.2cm}  i \in [1,m], \forall \hspace{.2cm} x_{1}, ... , x_{n} \in M, \neg R_{i}(x_{i}, ... ,x_{n})\\
then \hspace{.2cm} M  \\
else \hspace{.2cm} let \hspace{.2cm} x_{1}, ... , x_{n} \in M, let \hspace{.2cm} i \in [1,m] \hspace{.2cm} such \hspace{.2cm} that\\ \hspace{.2cm} R_{i}(x_{1}, ... ,x_{n}) \hspace{.2cm} in \hspace{.2cm}\\
\Gamma((R_{1}, A_{1}), ... ,(R_{m}, A_{m}))((M - {x_{1}, ... ,x_{n}}) +\\ A_{i}(x_{1}, ... ,x_{n}))
\end{split}\label{eq1}
\end{equation}~\\
[-10pt]

The pairs of functions $(R_{i}, A_{i})$ are applied to the multiset $(M)$ and specify the actions to be performed and their execution conditions. The execution of the pair $(R_{i}, A_{i})$ in $M$ results in replacing in $M$ a subset of elements $(x_{1}, ..., x_{n})$ such that $R_{i} (x_{1}, ..., x_{n})$ is true for elements of $A_{i} (x_{1}, ..., x_{n})$. If no element satisfies the reaction condition, the result of computing is the same initial $M$. Otherwise, the result is the multiset $M$ less the subset elements $(x_{1}, ..., x_{n})$ plus the subset specified by the reaction action $A_{i}(x_{1}, ..., x_{n})$, i.e., $((M - \{x_{1}, ..., x_{n}\}) + A_{i}(x_{1}, ..., x_{n}))$. 

It is important to note that while exists satisfied conditions for reactions executions, such reactions will be performed and consequently will transform the multiset. Therefore, if one or more reactions conditions are satisfied for some subsets of multiset simultaneously, the decision of which reaction will execute occurs in a nondeterministic way, since these reactions may be performed independently and simultaneously, which makes the Gamma computational model a naturally parallel environment \cite{artigo-paillard}.

For instance, considering the problem of choosing the smaller element in a some multiset. In the Gamma paradigm, this operation can be performed through a unique reaction, as depicted in Equation (\ref{eq:gamma_syntax}). We use the syntax of Gamma reactions based on the Gamma implementation provided by Juarez Muylaert \cite{artigo-paillard}.

\begin{equation}
\begin{split}	
R = replace (x,y)\\
by \hspace{0.2cm} x \hspace{1.9cm}   \\
where \hspace{0.2cm} x < y \hspace{0.7cm} 
\end{split}
\label{eq:gamma_syntax}
\end{equation}


Where the $R$ reaction compares two any elements $x$ and $y$, and return to the multiset only $x$, when $x < y$.

After the Gamma proposal in the 80s decade, some improvements needs have been verified, wich made the \textit{Structured Gamma} \cite{FRADET1998263} approach proposed in 1998. In this extended Gamma approach was introduced the concept of a structured multiset (making possible to represent data structures) and type checking at compile time. Some other Gamma extensions address operators composition, making possible the sequential and parallel reactions execution besides some semantic compositions \cite{10.1007/3-540-57502-2_57, 10.1007/BFb0039699, Métayer94higher-ordermultiset}.

Related to the Gamma paradigm implementations, we can quote some initial projects developed in the 90s decade as the \textit{Conection Machine} \cite{10.1007/3-540-55160-3_47} and \textit{MasPar} \cite{Huang1997} among others. Juarez Muylaert and Simon Gay proposed a sequential Gamma implementation \cite{artigo-paillard}. This project uses only one processor to perform reactions execution and does not allow the execution in a parallel hardware. This way, the sense of parallelism is obtained by the interchange between the reactions execution in a unique processor. The same authors, proposed a parallel Gamma implementation, where the reactions perform in a parallel hardware, with the communication between several processors is made through \textit{Message Passing Interface} (MPI) protocol. In 2015, another Gamma parallel implementation was proposed, now using a parallel hardware with support to \textit{Graphics Processing Unit} (GPU) \cite{deAlmeida:2016:GIG:2851613.2851719}. The Gamma paradigm can be applied to many application domains, for instance, Image Processing \cite{10.1007/3-540-45523-X_2} and data fusion for target tracking  \cite{7423190}, among others.

\section{Similarity Formalization} \label{sec:similarity_formalization}

Now we will present the main aspects needed to prove the equivalence between dataflow and Gamma computational models.


\subsection{Empirical Examples} \label{subsec:empirical_examples}


\subsubsection{Dataflow to Gamma} \label{subsubsec:dataflow_to_gamma}

Before showing details about the equivalence, focus of this paper, we present some simple examples to transform a dataflow graph into a Gamma code, and vice versa.

As our first example, consider the code bellow written in a high level language based on von Neumann paradigm:

\begin{footnotesize}
\begin{verbatim}
    int x = 1; 
    int y = 5;
    int k = 3;
    int j = 2;
    int m;
    m = (x + y) - (k * j);
\end{verbatim}
\end{footnotesize}

\begin{figure}[!ht]
\centering
\includegraphics[width=.43\textwidth]{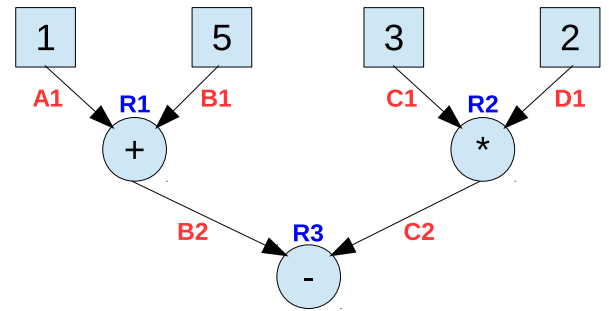}
\caption{Example 1 - dataflow graph}
\label{fig:figure_1}
\end{figure}

This program can be represented for the dataflow graph expressed in Figure \ref{fig:figure_1}, where all the vertices and edges were labeled in order to help in the conversion process.

Note that, in a dataflow paradigm, a program can be expressed in a directed graph, where the vertices and edges corresponds to operations and data, respectively. An vertex can have input operands, indicating that the instruction needs operands to perform, and output operands, where the instruction produces output data. A vertex is activated when all of yours input operands are available.

So, e. g., in the Figure \ref{fig:figure_1}, the subtraction operation, represented by the vertex $R3$, only can perform after data $B2$ and $C2$ have been produced by the operations (vertices) $R1$ and $R2$, respectively.  

To convert the graph expressed in the Figure \ref{fig:figure_1} in a Gamma code, all the vertices will be convert into reactions and the edges in multiset elements. Our initial multiset will be formed by the initial edges (output edges from the vertices represented by squares). As we need to save information about data and label (tagged information), our multiset elements will be represented by n-tuples of two elements. This way, the edge $A1$ correspond to the element $[1,A1]$, where the first information refers to the value of the edge and the second the label information.

This way, we have the follow initial multiset:

\begin{footnotesize}
\begin{verbatim}
    {[1, A1], [5, B1], [3, C1], [2, D1]}
\end{verbatim}
\end{footnotesize}

The transformation process convert all vertices into reactions that will manipulate and produce some data. For instance, the vertex $R1$ consumes the elements $[1, A1]$ and $[5, B1]$ producing the data $[1+5, B2]$, as follows:

\begin{footnotesize}
\begin{verbatim}
    R1 = replace [id1, `A1'], [id2, 'B1']
    by [id1 + id2, 'B2']
\end{verbatim}
\end{footnotesize}

Where $[id1, `A1']$ means a tuple tagged by $A1$ and having any value for the first field ($id1$). Note that there is no reaction condition expressed to the $R1$ reaction, once always there are two elements with label $A1$ and $B1$, this reaction occurs.  

This way, we can produce the follow Gamma code equivalent to the graph expressed in the Figure \ref{fig:figure_1}:

\begin{footnotesize}
\begin{verbatim}
    R1 = replace [id1, 'A1'], [id2, 'B1']
    by [id1 + id2, 'B2']

    R2 = replace [id1, 'C1'], [id2, 'D1']
    by [id1 * id2, 'C2']

    R3 = replace [id1, 'B2'], [id2, 'C2']
    by [id1 - id2, 'm']
\end{verbatim}
\end{footnotesize}

Note that in some implementations used as \cite{artigo-paillard}, the developer can introduce sequential and parallel operators related to the  reactions execution order, represented by ; and | respectively. In our examples we are considering only the parallel operator, that means that all reactions can run in parallel, i. e., $R1 | R2 | R3 | ... | Rn$.

Now consider a second example, represented for the following code:

\begin{footnotesize}
\begin{verbatim}
    For (i=z; i<0; i--)
        x = x + y;
\end{verbatim}
\end{footnotesize}

\begin{figure}[!ht]
\centering
\includegraphics[width=.46\textwidth]{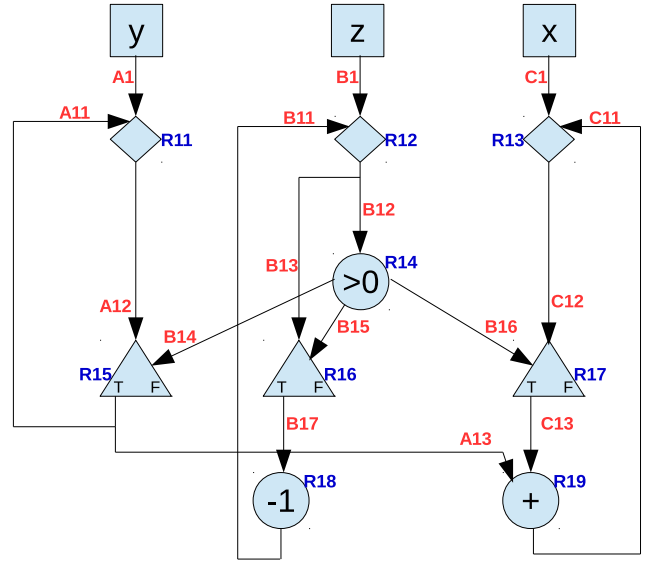}
\caption{Example 2 - dataflow graph}
\label{fig:figure_2}
\end{figure}

Likewise, the corresponding dataflow graph is presented in the Figure \ref{fig:figure_2}.

This example presents the concepts of loops and decision structures. For the decision structures, the dataflow graph present the operator \textit{Steer} (represented in the Figure \ref{fig:figure_2} by triangles). This operator receive two input operands. The first one is the data value and the second a Boolean control signal. If the control signal is true, the true output receive the input data value, otherwise, the false output transmit the data value. As we can see in the Figure \ref{fig:figure_2}, the graph present three \textit{Steer}, labeled by $R15$, $R16$ and $R17$. All of control signals received by these \textit{Steer} are produced through the comparison with zero, expressed in the vertex $R14$.

Consider the \textit{Steer} $R16$, in the Figure \ref{fig:figure_2}. It receives two input operands ($B13$ and $B15$) and produces only the true output ($B17$). The correspondent Gamma code can be represented by:

\begin{footnotesize}
\begin{verbatim}
    R16 = replace [id1,'B13',v], [id2,'B15',v]
    by [id1,'B17',v]
    if id2 == 1
    by 0
    else
\end{verbatim}
\end{footnotesize}

Here, case the Boolean control signal ($B15$) was true ($id2 == 1$), the elements $B13$ and $B15$ will be replaced by the element $B17$ which will contain the value of $B13$ data. Otherwise, these two elements will be excluded from the multiset and no other element will be inserted (else clause).

Another special kind of vertex is the \textit{Inctag}, represented in the dataflow graph by lozenges $R11$, $R12$ and $R13$. This operator is responsible for increment the iteration label of each data (operand), identifying data of different iterations. This way, a operation only can occurs with data of the same iteration label. For this reason, in this example, the data will be represented by a n-tuple composed by three elements: data, edge label and iteration label. So, the initial multiset for the graph expressed by the Figure \ref{fig:figure_2}, is: 

\begin{footnotesize}
\begin{verbatim}
    {{y, A1, 0}, {z, B1, 0}, {x, C1,0}}
\end{verbatim}
\end{footnotesize}

Note that, for the initial multiset elements, all of iteration label are equals to zero. This value will be increment, in each iteration, as an effect of the \textit{Inctag} operation. The concepts of \textit{Steer} and \textit{Inctag} were presented in \cite{5645392}.

Now consider an example of \textit{Inctag}, the vertex $R11$. It receives only one input operand ($A1$ for the first iteration or $A11$ for the others) and produces also only one output ($A12$). The correspondent Gamma code can be represented by:

\begin{footnotesize}
\begin{verbatim}
    R11 = replace [id1,x,v]
    by [id1,'A12',v+1]
    if (x=='A1') or (x=='A11')
\end{verbatim}
\end{footnotesize}

Note that this reaction only increments the iteration label and change (transform) the edge label of the data.

Now, according the transformations presented in the first example, and the observations about the \textit{Steer} and \textit{Inctag} operators, the Gamma code correspondent to the dataflow graph showed in the Figure \ref{fig:figure_2}, can be expressed through nine reactions, as follow:

\begin{footnotesize}
\begin{verbatim}
    R11 = replace [id1,x,v]
    by [id1,'A12',v+1]
    if (x=='A1') or (x=='A11')

    R12 = replace [id1,x,v]
    by [id1,'B12',v+1], [id1,'B13',v+1]
    if (x=='B1') or (x=='B11')

    R13 = replace [id1,x,v]
    by [id1,'C12',v+1]
    if (x=='C1') or (x=='C11')

    R14 = replace [id1, 'B12', v]
    by [1,'B14',v], [1,'B15',v], [1,'B16',v]
    If id1 > 0
    by [0,'B14',v], [0,'B15',v], [0,'B16',v]
    else

    R15 = replace [id1,'A12',v], [id2,'B14',v]
    by [id1,'A11',v], [id1,'A13',v]
    If id2 == 1
    by 0
    else
	
    R16 = replace [id1,'B13',v], [id2,'B15',v]
    by [id1,'B17',v]
    If id2 == 1
    by 0
    else	

    R17 = replace [id1,'C12',v], [id2,'B16',v]
    by [id1,'C13',v]
    If id2 == 1
    by 0
    else

    R18 = replace [id1,'B17',v]
    by [id1 - 1,'B11',v]

    R19 = replace [id1,'A13',v], [id2,'C13',v]
    by [id1+id2,'C11',v]
	
\end{verbatim}
\end{footnotesize}


\subsubsection{Gamma to dataflow} \label{subsubsec:gamma_to_dataflow}

With respect to convert a gamma code into a dataflow graph, the basic idea is to transform each reaction in a vertex and each data manipulated by this reaction in edges.

Considering again the first gamma code example presented in section \ref{subsubsec:dataflow_to_gamma}, related to the \ref{fig:figure_1}. In this example, the Gamma code was composed by three reactions, $R1$, $R2$ and $R3$. 

From the reaction $R1$, will be created a vertex, called $R1$, corresponding to a sum operation, described by the clause "by":

\begin{footnotesize}
\begin{verbatim}
    R1 = replace [id1, 'A1'], [id2, 'B1']
    by [id1 + id2, 'B2']
\end{verbatim}
\end{footnotesize}~\

This $R1$ vertex has two input operands ($[id1, 'A1']$ and $[id2, 'B1']$) and produce only one operand ($[id1 + id2, 'B2']$). That way, in a dataflow graph, the vertex $R1$ will have two inputs operands $A1$ and $B1$ and produce one output operand, $B2$. Similar process can be performed for the reactions $R2$ and $R3$. Finally, the initial multiset composed by the elements $[1, A1]$, $[5, B1]$, $[3, C1]$ and $[2, D1]$, will give rise to the initial vertices and edges, represented by square vertices. Thus, we can reproduce the same dataflow graph of the Figure \ref{fig:figure_1} from the three reactions mentioned.

Similar reasoning can be executed over the second example of the section \ref{subsubsec:dataflow_to_gamma}, composed by nine reactions. The reactions $R11$, $R12$ and $R13$ refers to \textit{Inctag} operator. Consider the $R11$ reaction. Here, we have only one input operator (as mentioned by the clause "replace"), however, this operand can have the edge labels equals to $A1$ or $A11$ (according reaction condition $if(x=='A1') or (x=='A11')$). The \textit{Inctag} operation can be identified by the increment of the iteration label field ($by [id1, 'A12', v+1]$). This way, this kind of vertex will receive one input operand ($A1$ or $A11$) and will produce only one output operand, $A12$, and will be represented as a lozenge. 

In relation to the \textit{Steer}, represented by reactions $R15$, $R16$, and $R17$, all of them consumes two multiset elements and produces elements related to the true test condition (expressed in the clauses $by$ and $if$). In other words, in this example, only the true output will produce elements. So the \textit{Steer} can be identified by always compare two elements (value and Boolean control signal) and have conditional tests for true and false clauses. Thus, this kind of reaction can be convert into a vertex represented by a triangle. Considering the $R7$ reaction, the input operands will be $C12$ and $B16$ ($replace [id1,'C12',v], [id2,'B16',v]$), and only the true output will be provided ($by [id1,'C13',v]$), creating the edge labeled as $C13$, case $id2$ true. Thus, applying the same process of the first example, we can reproduce the dataflow graph presented in the Figure \ref{fig:figure_2}.


\subsubsection{Reductions} \label{subsubsec:Reductions}

The number of Gamma reactions presented in the first and second examples of the section \ref{subsubsec:dataflow_to_gamma}, can be reduced to decrease the final number of reactions. This fact will be directly affect the granularity of operations in both Gamma and dataflow models. So, some reductions or expansions can be performed.

Considering the Gamma code composed by reactions $R1$, $R2$ and $R3$, referred to the conversion of the Figure \ref{fig:figure_1}, this reactions can be replaced by only one reaction, as follow:

\begin{footnotesize}
\begin{verbatim}
    Rd1 = replace [in1,'A1'], [id2,'B1'], 
    [id3,'C1'], [id4,'D1']
    by [(id1+id2)-(id3*id4),'m']
\end{verbatim}
\end{footnotesize}

Note that with this reduced code, the opportunity of explore the parallelism of reactions decrease, once that this reaction only will occurs, when the all of operands are chosen in the order expected for this reaction. In other words, the chance of the reactions condition occurs can decrease.

Also the gamma code of the second example (referred to the Figure \ref{fig:figure_2}) can be reduced. In this case, we managed to reach a total of six reactions, as follow:

\begin{footnotesize}
\begin{verbatim}
    Rd11 = replace [id1,x,v]
    by [id1,'A12',v+1]
    If (x=='A1') or (x=='A11')

    Rd12 = replace [id1,x,v]
    by [id1,'B14',v+1], [id1,'B12',v+1], 
    [id1,'B16',v+1] 
    If (x=='B1') or (x=='B11')

    Rd13 = replace [id1,x,v]
    by [id1,'C12',v+1]
    If (x=='C1') or (x=='C11')

    Rd14 = replace [id1,'A12',v], [id2,'B14',v]
    by [id1,'A11',v], [id1,'A13',v]
    If id2 > 0
    by 0 
    else
	
    Rd15 = replace [id1,'B12',v]
    by [id1 - 1,'B11',v]
    If id1 > 0
    by 0
    else

    Rd16 = replace [id1,'A13',v], [id2,'B16',v], 
    [id3,'C12',v] 
    by [id1 + id3,'C11',v]
    If id2 > 0
    by 0 
    else
\end{verbatim}
\end{footnotesize}

\subsection{Conversion Algorithm} \label{subsec:conversion_algorithm}

In this section, we present the algorithms used for the transformations from a dataflow graph to a Gamma code and from a Gamma code to a dataflow graph, presented in section \ref{subsec:empirical_examples}.

\begin{figure}[ht]
    \centering  
    \includegraphics[width=\linewidth]{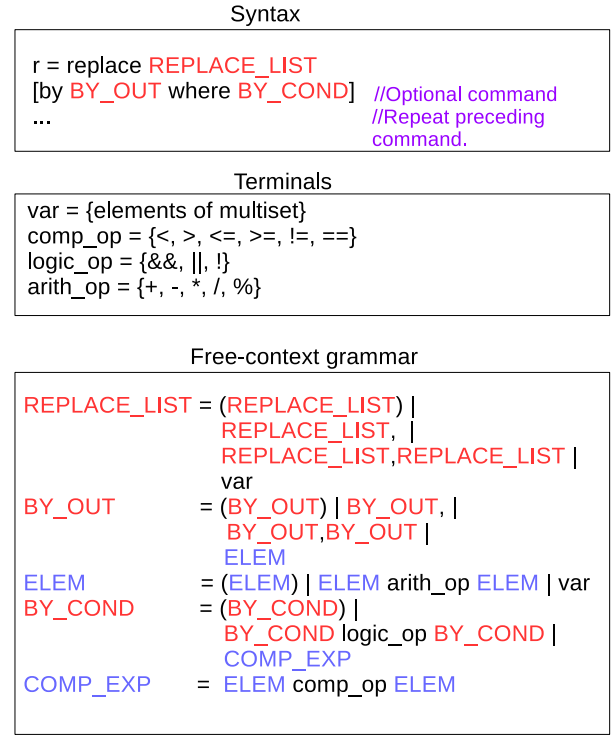}
    \caption{Free-context grammar of Gamma syntax.}
    \label{fig:gamma_syntax}
\end{figure}

We generalize Gamma syntax presented in Equation (\ref{eq:gamma_syntax}) into free-context grammar notation as shown in Figure \ref{fig:gamma_syntax}. Basically the syntax is composed of two parts: \textbf{replace list} that describe the number of elements to initialize the reaction and \textbf{by list} that specifies the produces elements in \textbf{by output} controlled by conditions in \textbf{by condition}. By using free-context grammar notation, Gamma code can be trivially created by reading data structures that keep the \textit{replace list} and \textit{by list} part.

The procedure to generate a Gamma program from dataflow graph is detailed in Algorithm \ref{alg:dataflow2gamma}. Initially, it generates a label for each node in dataflow graph (lines 3-6). As discussed in Section \ref{subsubsec:dataflow_to_gamma}, to support \textit{Inctag} instructions each element of multiset has to be a triplet \textit{[value, label, tag]}. Initial multiset $M$ is created by root nodes at line 9, since they have no input operands. The other nodes add parameters in the \textit{replace list} $R_L$ and the \textit{by list} $B_L$ where each entry contains the output values $B_V$ and conditions $B_C$. \textit{Steer} nodes produces two entries in \textit{by list} associated to each path that the output operands can be sent (true $t$ or false $f$ port) controlled by boolean operand $x_1$. At lines 21-22, \textit{Inctag} nodes only inclement the tag from input elements. Arithmetic and comparison operator nodes produces their operations in \textit{by list} replicating output elements with label for all output nodes in the dataflow graph.

\begin{algorithm}[!htbp]
  
    \SetAlgoLined
  
    \KwIn{Dataflow graph $D(I, E)$, $values$ array storing value for each node}
    \KwOut{Gamma program $G(R, M)$ corresponding to $D(I, E)$ }
    
    $label\gets []$; $l\gets 0$\;
    $R\gets \varnothing$; $M\gets \varnothing$\;  
    
    \ForEach{instruction $i \in I$}{
        $label[i]\gets l$\;
        $l\gets l + 1$\;
    }
    
    \ForEach{instruction $i \in I$}{
        \uIf{$i$ is root}{
            $M\gets \{[value(i), label[i], 0]\}$\;
        }\Else{
            $R_L\gets \varnothing$;\tcp{Replace list}\
            $B_L\gets \varnothing$;\tcp{By condition list}\
            \uIf{$i$ is Steer $st$ with input $s$ and output ports $t$ and $f$ }{
                $R_L\gets \{[x_{0}, label[s], tag], [x_{1}, label[st], tag]\}$\;
                $B_V1\gets \{[x_{0}, label[t], tag]\}$\;
                $B_C1\gets \{(x_{1} == 1)\}$\;
                $B_V2\gets \{[x_{0}, label[f], tag]\}$\;
                $B_C2\gets \{(x_{1} == 0)\}$\;
                $B_L\gets \{(B_V1, B_C1), (B_V2, B_C2))\}$\;
            }\uElseIf{$i$ is Inctag $it$ with input $s$ and output $o$}{
                $R_L\gets \{[x_{0}, label[it], tag]\}$\;
                $B_L\gets \{[x_{0}, label[o], tag + 1]\}$\;
            }\uElseIf{$i$ is comparison operator $op$ with inputs $s1$ and $s2$}{
                $R_L\gets \{[x_{0}, label[s1], tag], [x_{1}, label[s2], tag]\}$\;
                
                \ForEach{output $o$ from $i$}{            
                    $B_L\gets \{[1, label[o], tag], (x_{0}$ $op$ $x_{1})\}$\;
                    $B_L\gets \{[0, label[o], tag], !(x_{0}$ $op$ $x_{1})\}$\;
                }
            }\ElseIf{$i$ is arithmetic operator $op$ with inputs $s1$ and $s2$}{
                $R_L\gets \{[x_{0}, label[s1], tag], [x_{1}, label[s2], tag]\}$\;
            
                \ForEach{output $o$ from $i$}{            
                	$B_L\gets \{[x_{0}$ $op$ $x_{1}, label[o], tag]\}$\;
                }
            }
            
            $R\gets \{(R_L, B_L)\}$\;
        }
    }

    \caption{Algorithm for converting from dataflow graph to reaction set in Gamma.}
    \label{alg:dataflow2gamma}    
\end{algorithm}

The procedure to convert from Gamma code to dataflow graph is divided in two steps: ($1$) generate a dataflow graph for each reaction and ($2$) map the multiset elements along the dataflow graphs produced in step $1$.
Step $1$ is presented in Algorithm \ref{alg:gamma2dataflow}. Considering each reaction is associated to a dataflow graph, the root nodes are obtained by elements in \textit{replace list} $R_L$ at lines 2-4. If \textit{by list} $B_L$ has no condition expression, then arithmetic nodes and the edges, connecting each input element from \textit{replace list} to arithmetic operators, are created at lines 18-21. Otherwise, \textit{Steer} nodes are generated with related comparison nodes and their true port are linked to the arithmetic nodes at lines 13-16. Note that, only analyzing reaction syntax does not provide enough information to produce \textit{Inctag} nodes. Loops are implicitly describes in Gamma program with undetermined number of iterations.

\begin{figure}[ht]
    \centering  
    \includegraphics[width=\linewidth]{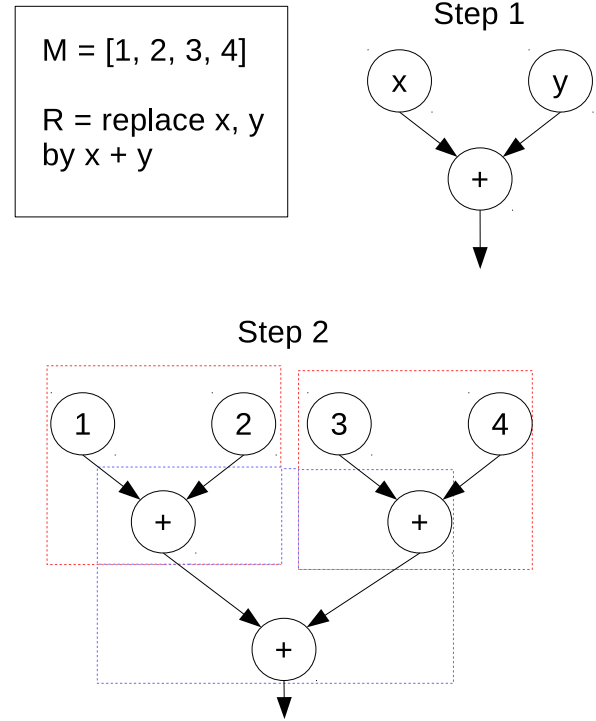}
    \caption{Gamma to dataflow graph example.}
    \label{fig:gamma_dataflow}
\end{figure}

As elements of the multiset are chosen to be processed by a reaction during run-time, to map multiset elements in the Step $2$ is need to combine all elements of initial multiset $M$ to the root nodes of the dataflow graphs. This process requires to replicate dataflow graphs to fit the whole multiset. The produced elements have to be connected to the dataflow graph until the reactions finishing their processing. An example is shown in Figure \ref{fig:gamma_dataflow}, where dataflow graph generated from reaction $R$ is instanced $3$ times to connect all elements of the multiset.
The algorithm that efficiently maps elements to dataflow graph is complex and it is beyond the scope of this work.

\begin{algorithm}
  
    \SetAlgoLined
  
    \KwIn{Reaction $R(R_L, B_L)$ with replace list $R_L$ and by list $B_L$ }
    \KwOut{Dataflow graph $D(I, E)$ }
    
    $I\gets \varnothing$; $E\gets \varnothing$\;  
    
    \ForEach{input element $e \in R_L$}{
        $I\gets$ node $i$; $value[i] \gets e$\;
    }
    
    \ForEach{by command $(B_V, B_C) \in B_L$}{
        \uIf{$B_C$ is not empty}{
            \ForEach{comparison expression $exp \in B_C$}{
                $I\gets$ comparison node $c, \forall$ comparison operator $op \in exp$\;
                $E\gets (e, c), \forall e \in R_L$ used as input to comparison operator $op \in exp$\;
                $I\gets$ Steer node $st, \forall e \in R_L$ affected by result of comparison operator $op\in exp$\;
                $E\gets (e, st), \forall e \in R_L$ affected by result of operator $op \in exp$ related to Steer node $st$\;
            }
            
            \ForEach{arithmetic expression $exp \in B_V$}{
                $I\gets$ arithmetic node $a, \forall$ arithmetic operator $op \in exp$\;
                $E\gets (se.true, a), \forall$ Steer node $se$ related with $e \in R_L$ used as input to arithmetic operator $op \in exp$\;
            }
        }\Else{
        
            \ForEach{arithmetic expression $exp \in B_V$}{
                $I\gets$ arithmetic node $a, \forall$ arithmetic operator $op \in exp$\;
                $E\gets (e, a), \forall e \in R_L$ used as input to arithmetic operator $op \in exp$\;
            }
        }
    }

    \caption{Algorithm for converting from reaction to dataflow graph.}
    \label{alg:gamma2dataflow}    
\end{algorithm}



\subsection{Sketch of Proof} \label{subsec:sketch_proof}
We can show that Algorithm \ref{alg:dataflow2gamma} produces a set of reactions for gamma equivalent to the dataflow graph provided as input. The fundamental property of dataflow, i.e., an operation \emph{op} is only triggered when its operands are ready, is guaranteed by the creation of equivalent reaction in line 29. Notice that this reaction also takes into account the tag, corresponding to the iteration in case of loops in a dynamic dataflow graph. 

In lines 21 and 22 we maintain the tags corresponding to loops iterations by adding a reaction that replaces (in the multiset) the input operand of an \textit{Inctag} with the same value, incrementing the value operand's tag. This produces in the final gamma program the same effect of the loop in the dataflow graph. Finally, lines 13-19 provide the effect of \textit{Steer} instructions by adding the conditions of such instructions to the corresponding reactions.


\section{Concluding remarks} \label{sec:concluding_remarks}

Dynamic dataflow and Gamma paradigms are emerging as computational options to meet the recent challenges for naturally parallel computation. In this context, the equivalence between both computational models, Gamma and dynamic dataflow, is exposed here, for the first time. This equivalence has the potential of providing several extra benefits. For instance, enable the analysis of trace reuse in a dataflow graph through a Gamma code and to perform Gamma reactions correspondent to a dataflow code in a distributed multiset environment.

This paper presented the motivation and the objectives of proving the equivalence between these two apparently very different models. One can affirm that the expression of parallelism in these two environments occur in a very natural way. The basic concepts related to Gamma and dynamic dataflow were presented, and the main related works were addressed throughout the text. Some basic, though representative examples, were provided, and the transformation details between the two models were discussed.

This work also contributes to program development versatility, since the developer may choose to express program specifications in the two distinct computational models, both containing powerful and natural mechanisms for parallelism exploration. Given the possibility of transformation between the two models, benefits from both sides can be widely exploited, since, for instance, a programmer with mathematical background could define his programs in Gamma, while benefiting from a dynamic dataflow execution environment.

Expliciting the transformations needed to complete the conversion algorithm of a Gamma code into a dynamic dataflow graph is subject for future work. Such transformations are related to identify kinds of dataflow nodes (\textit{steer}, \textit{inctag}, etc) via the analysis of the behavior of Gamma reactions. Also left for future exploration is the implementation of the transformation algorithms presented in section \ref{subsec:conversion_algorithm}. Another interesting research thread left unexplored is the exploitation of interest-based communication protocols, like \textit{Information Centric Networks} (ICN), in the development of an Internet of Things (IoT) environment, via the implementation of Gamma distributed multisets. 

\section*{Acknowledgment}

This study was financed in part by the Coordena\c{c}\~ao de Aperfei\c{c}oamento de Pessoal de N\'ivel Superior - Brasil (CAPES) - Finance Code 001.

The authors would like to thank also the Brazilian Navy Research Institute (IPqM), CNPq and FAPERJ.

\bibliographystyle{IEEEtran}
\bibliography{references}

\end{document}